\documentclass{article}

\PassOptionsToPackage{numbers}{natbib}
 \usepackage[creativeai, final]{neurips_2025}

\usepackage[utf8]{inputenc} % allow utf-8 input
\usepackage[T1]{fontenc}    % use 8-bit T1 fonts
\usepackage{hyperref}       % hyperlinks
\usepackage{url}            % simple URL typesetting
\usepackage{booktabs}       % professional-quality tables
\usepackage{amsfonts}       % blackboard math symbols
\usepackage{nicefrac}       % compact symbols for 1/2, etc.
\usepackage{microtype}      % microtypography
\usepackage{xcolor}         % colors

\usepackage{url}
\usepackage{graphicx}
\usepackage{color}
\usepackage{bbm}
\usepackage{amsfonts, amsmath, amssymb, amsthm}
\usepackage{algorithm}
\usepackage{algpseudocode}
\usepackage{graphicx}
\usepackage{cleveref}

\usepackage{wrapfig}

\usepackage{booktabs}

\newcommand\ourmethod{{\textrm{ProGress}}}
\title{ProGress: Structured Music Generation via Graph Diffusion and Hierarchical Music Analysis}

\author{%
  Stephen Ni-Hahn\thanks{Equal contribution} \\
  Duke University\\
  \texttt{stephen.hahn@duke.edu} \\
  \And
  Chao Péter Yang$^*$ \\
  Duke University \\
  \texttt{peter.yang@duke.edu} \\
  \AND
  Mingchen Ma \\
  Duke University \\
  \And
  Cynthia Rudin \\
  Duke University \\
  \And
  Simon Mak \\
  Duke University \\
  \And
  Yue Jiang \\
  Duke University \\
}

\begin{document}

\maketitle

\begin{abstract}
Artificial Intelligence (AI) for music generation is undergoing rapid developments, with recent symbolic models leveraging sophisticated deep learning and diffusion model algorithms. One drawback with existing models is that they lack structural cohesion, particularly on harmonic-melodic structure. Furthermore, such existing models are largely ``black-box'' in nature and are not musically interpretable. This paper addresses these limitations via a novel generative music framework that incorporates concepts of Schenkerian analysis (SchA) in concert with a diffusion modeling framework. This framework, which we call ProGress (\underline{Pro}longation-enhanced Di\underline{Gress}), adapts state-of-the-art deep models for discrete diffusion (in particular, the DiGress model of Vignac et al., 2023) for interpretable and structured music generation. Concretely, our contributions include 1) novel adaptations of the DiGress model for music generation, 2) a novel SchA-inspired phrase fusion methodology, and 3) a framework allowing users to control various aspects of the generation process to create coherent musical compositions. Results from human experiments suggest superior performance to existing state-of-the-art methods.
\end{abstract}

\section{Introduction}
\label{sec:intro}

Music technology is expanding at a rapid pace with increasing focus on artificial intelligence (AI) \citep{lee2024ai}. Many generative audio AI companies and models have arisen recently, targeting domains such as video background music \citep{beatoven}, responsive video game music \citep{infinitealbum}, sleep and focus aids \citep{endel}, and language-guided generation \citep{suno, stableaudio, riffusion, agostinelli2023musiclm, copet2024simple}. In particular, AI for symbolic music -- music that can be written in a score -- is a newly important topic in academic spheres \citep{roberts2018hierarchical, wu2019hierarchical, mittal2021symbolic, 
hahn_schenkcomposer, plasser2023discrete, luo2024bandcontrolnet, jonason2024symplex}. 

%Perhaps the most widely-known among these models, Suno and Stable Audio, rely on vasts amounts of copyrighted data to achieve high-quality outputs, leading to ethical concerns involving authorship, credit, and lack of licensing \citep{newton_rex2024, newton_rex2025}. MAKE DISTINCTION BETWEEN AUDIO AND SYMBOLIC MUSIC.

One major concern with current music generation AI is a lack of music-theoretical awareness. Most existing models target the learning of music-theoretical principles in an implicit fashion by processing massive amounts of (often unethically sourced) data \citep{newton_rex2024}, primarily using massive models with hundreds of millions of parameters. Due to this reliance on training data without guidance from musical principles, such models fail to capture true musical structure, resulting in generated music that is incoherent, difficult to follow, and that sounds more like a ``stream of consciousness.'' Several models have incorporated structure through musical form or meter, constraining music to a verse-chorus structure or encoding notes grouped by measure, e.g., \citep{roberts2018hierarchical, wu2019hierarchical, zou2022melons, suno}. However, such approaches do not account for the more detailed and complex voice-leading structure that is necessary for defining a musical style. 

Looking to build more organically-structured music models that are guided by domain knowledge, recent promising work has incorporated features of \textit{Schenkerian analysis} (SchA) and music-theoretical concepts within learning model algorithms and architectures \citep{hahn_schenkcomposer, ni2024new, ours2024cluster, ours2025schenkerlink}. Along this vein, this paper introduces a novel generative symbolic music framework that incorporates aspects of hierarchical music theory in concert with deep learning. Our framework, which we call ProGress (\underline{Pro}longation-enhanced Di\underline{Gress}), builds on state-of-the-art deep models for discrete graph diffusion \citep{vignac2023digressdiscretedenoisingdiffusion, huang2024symbolicmusicgenerationnondifferentiable} with a careful integration of well-established music composition principles from SchA. In doing so, our framework allows users to control various aspects of the generation process in an interpretable manner to create novel, coherent, and musically pleasing compositions, even with highly limited training data. Concretely, our contributions include 1) novel adaptations of the DiGress model for music generation, 2) a novel SchA-inspired phrase fusion methodology, and 3) a framework allowing users to control various aspects of the generation process to create structurally coherent music. We emphasize that our model uses orders of magnitude fewer parameters than current state-of-the-art competitors, while producing \textit{superior} generated music as evaluated by blinded human experiments.

The paper is structured as follows. Section \ref{sec:scha_background} provides background information on SchA. Section \ref{sec:methodology} presents the proposed ProGress modeling framework. Section \ref{sec:experiments} discusses our experiments including a blinded human experiment, ablation studies, and genre transferability.

\section{Background on Schenkerian Analysis}
\label{sec:scha_background}

Schenkerian analysis (SchA) is a powerful tool for representing music's hierarchical harmonic-melodic structure, showing how harmonies are ``unfolded'' through time in the form of melodies \citep{cadwallader1998analysis, schenker2001free}. Vitally, SchA reveals recursive patterns in music at various levels of structure; the musical foreground (music as it is written in the score) hosts similar harmonic-melodic progressions to events in the musical middleground and background. 
While SchA was originally designed for western classical music of the common practice era (ca$.$ 1600-1900), it has been adapted for analyzing music from all over the world, from Chinese opera to Ghanaian folk music \citep{stock1993application}, and over broad time periods and styles, from medieval polyphony \citep{salzer1967medieval} to modern rock \citep{nobile2014}. 
\begin{wrapfigure}{r}{0.60\textwidth}
    \centering
    \includegraphics[width=\linewidth]{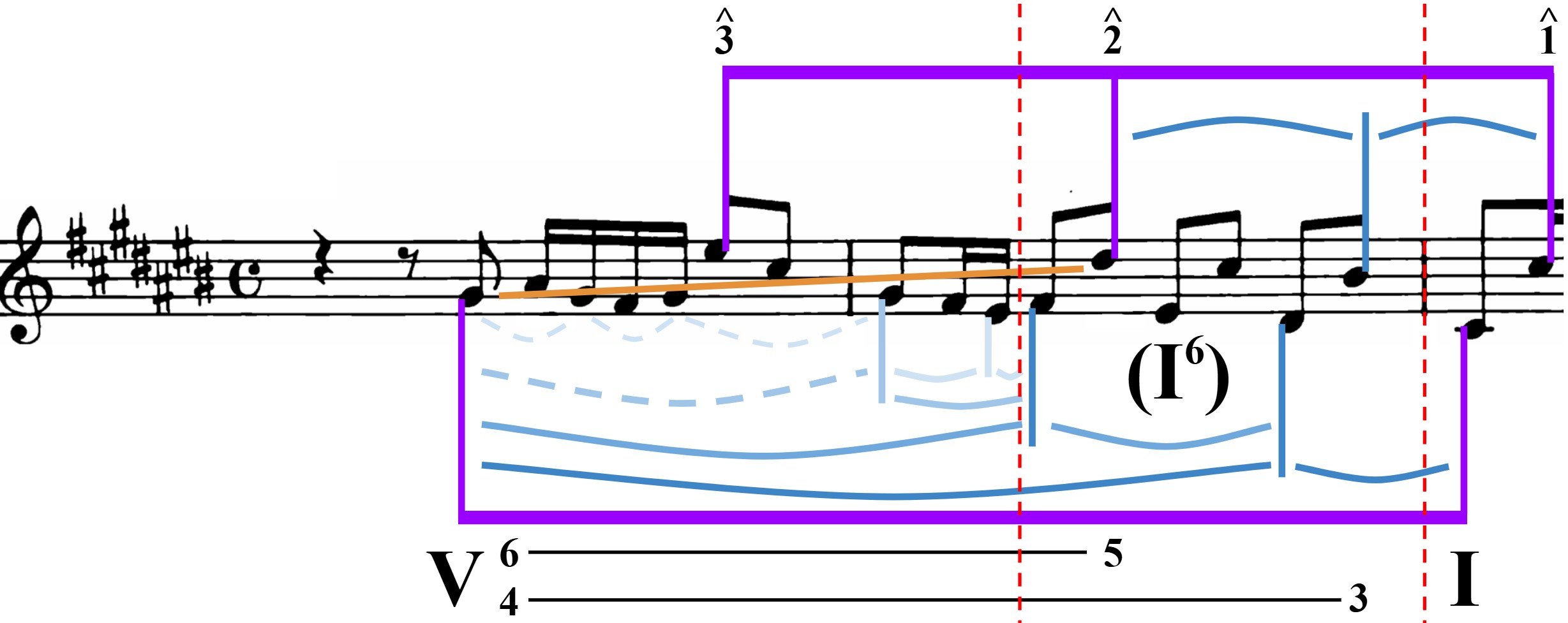}
    \caption{Example SchA of J.S. Bach's C$\sharp$ major fugue subject from \textit{Das Wohltemperierte Klavier I}.}
    \label{fig:scha_example}
\end{wrapfigure}

Figure \ref{fig:scha_example} provides an example of the first author's analysis of Bach's C$\sharp$ major fugue subject from \textit{Das Wohltemperierte Klavier I}. Here, we represent more foreground structures with lighter blue stems and slurs, while deeper middleground structures are represented with darker blue. The background structure is represented with purple. The background upper voice outlines a 3rd progression (E$\sharp$-D$\sharp$-C$\sharp$ or $\hat{3}$-$\hat{2}$-$\hat{1}$), which is a common cadential melodic pattern in tonal music. The background harmonic structure is described in Roman numerals at the bottom with red dotted lines to separate major harmonic shifts. The first measure outlines a cadential V, while measure 2 unfolds the resolution of the 6th and 4th to the tones of a dominant (G$\sharp$) harmony, which resolves to tonic I (C$\sharp$) in measure 3. The orange line connecting the bass G$\sharp$ in measure 1 to the treble D$\sharp$ in measure 2 clarifies that they are part of the same harmony in the background structure, separated by a relatively large span of time. 

Note that the parenthetical I$^6$ in measure 2 is understood as a foreground passing harmony, \textit{prolonging} the dominant V harmony that surrounds it. Prolongation (the inspiration for our model's name) refers to the phenomenon where certain notes or harmonies are ``in control'' at deeper levels of structure. While this example is relatively short, similar recursive prolongational relationships can span entire sections, movements, or even opuses.

By incorporating such harmonic-melodic structure in the generative process, music generative models can connect broader structures and produce more cohesive compositions, thus addressing a key limitation of many existing AI-based models. \textit{ProGress}, presented next, aims to do this within a carefully-structured deep learning framework.

%Recently, \citep{ours2025schenkerlink} developed a GNN-based model, named SchenkerLink, to estimate the detailed hierarchical melodic-harmonic structure of music. SchenkerLink poses the problem of SchA as a graph link prediction task. Edges are sampled in a music-theoretically structured way based on the predicted probabilities distributed over all musically plausible pairwise node connections. While the SchenkerLink model is designed to estimate structural edges from the full musical score, there is evidence that SchA models can make meaningful structural determinations based on rhythmic features alone \citep{ours2024cluster}. We therefore make use of a rhythmic SchenkerLink model (without pitch features) within our ProGress model.

%\citep{rgcn-17} proposes R-GCN, which proposes the heterogeneous analog to the standard graph convolution operation of \citep{kipf2016semi}. To handle directed graphs, we use the framework given by \citep{dir-gcn-23}, where additional parameters are introduced for each edge to encode directionality.

\section{Methodology}
\label{sec:methodology}

\begin{figure*}
    \centering
    \includegraphics[width=1.0\linewidth]{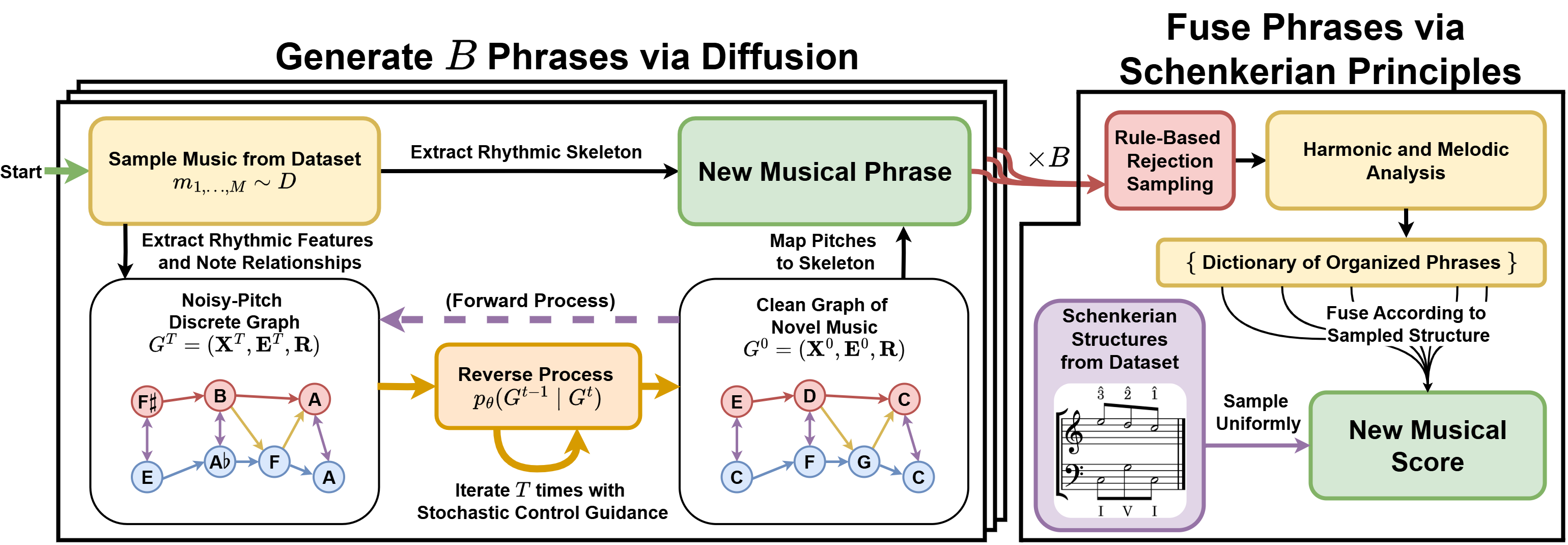}
    \vspace{-0.3cm}
    \caption{Overview of the phrase generation process. On the left half, B phrases are generated via diffusion, and on the right, phrases are fused together according to music theoretical principles and structures. In the generation stage, starting with the yellow block, we sample a phrase from our dataset $D$ and extract rhythmic relationships to build a heterogeneous, discrete graph $G^T$. This discrete graph is iteratively passed through a denoising model $p_\theta$ to determine the notes for a novel piece of music. Finally, the inferred notes are mapped back to the rhythmic skeleton of the sampled phrase. This process is repeated $B$ times to generate $B$ phrases. For the fusion stage, phrases are first analyzed and organized based on harmonic and structural features. Based on user-defined rules, certain phrases are rejected. Phrases are then fused according to a sampled Schenkerian structure as described in Section \ref{subsec:inference}.}
    \label{fig:model_overview}
    \vspace{-0.3cm}
\end{figure*}

We now describe the proposed ProGress music generation framework via discrete graph diffusion and prolongation-enhanced phrase fusion; Figure \ref{fig:model_overview} visualizes its workflow. ProGress first extends a state-of-the-art diffusion model to generate a broad library of diverse musical phrases. After passing such phrases through rule-based rejection sampling, \ourmethod{} analyzes and organizes harmonic and melodic qualities of accepted phrases. Based on a sampled Schenkerian structure, individual phrases are then fused together into a structured score using music-theoretical principles. Section \ref{subsec:rhythmic_sampling} describes how rhythmic information is extracted from a music dataset. Section \ref{subsec:diffusion_modeling} describes the musical representation, implementation details, and adaptations required for the employed diffusion model. Section \ref{subsec:inference} describes our phrase fusion methodology in finer detail.

\subsection{Rhythmic Sampling} 
\label{subsec:rhythmic_sampling}

First, we sample from a musical dataset $D$ or user input to determine a rhythmic framework as the backbone of our new music. There are numerous ways of performing this sampling to generate structured music. The simplest is to sample entire phrases from the dataset and extract their rhythm. Another is to uniformly sample various measures $m_1,\dots,m_M$ from $D$ and combine them into one phrase. If the latter approach is employed, it is useful to sample measures that end phrases separately, as cadential motions are often more carefully constructed. Phrases can further be combined and varied according to common patterns in the desired genre. For our figures and experiments, we focus on the case where rhythmic samples consist of entire phrases. 

\subsection{Discrete Graph Diffusion Modeling for Music}
\label{subsec:diffusion_modeling}

Next, we generalize the Discrete Graph Denoising Diffusion Model (DiGress) in \citep{vignac2023digressdiscretedenoisingdiffusion} for musical phrase generation (see Appendix A for background and details on DiGress). The goal of such a model is to build a library of diverse musical phrases (i.e., set of pitches) given a rhythmic framework. For this model, graphs are defined by categorical node and edge attributes.

%In contrast to SchenkerLink, there is only one edge type, but each edge has categorical features associated with it. This subtle distinction affects how the denoising model handles message passing and limits the number of edges to $n*n$ rather than SchenkerLink's $n*n*m$. Furthermore, nodes only belong to a single category. 

%Given a set of node categories $\mathcal{X}$ and edge categories $\mathcal{E}$, a graph $G = (\mathbf{X}, \mathbf{E})$ is comprised of node embedding matrix $\mathbf{X} \in \{0,1\}^{n\times |\mathcal{X}|}$, where each row is a one-hot encoding $\mathbf{x}_i \in \{0,1\}^{|\mathcal{X}|}$ for graph nodes $i = 1,\dots,n$, and edge embedding tensor $\mathbf{E} \in \{0,1\}^{n\times n\times |\mathcal{E}|}$, which describes each edge $\mathbf{e}_{i,j} \in \{0,1\}^{|\mathcal{E}|}$ from node $i$ to node $j$ as a one-hot encoding. Note that the absence of an edge or node is represented as a particular class. Thus, all $\mathbf{x}_i$ and $\mathbf{e}_{i,j}$ are non-empty and have one entry indicating its category.

\textbf{\textit{Node and Edge Categories:}} The nodes in our DiGress model each belong to a category from $\mathcal{X} = \{\hat{1}, \sharp\hat{1}, \flat\hat{2}, \hat{2},\dots,\sharp\hat{6},\flat\hat{7},\hat{7}, \text{rest}\}$, representing the \textit{global scale degree} of a musical note, i.e, the note's place within the context of a piece's home key. The node category is our primary interest for inference, as it transforms the rhythmic framework into theoretical music with pitch classes. Edges each belong to a category from $\mathcal{E} = \{\text{forward}, \text{treble-voice},\text{bass-voice},\text{onset},\text{sustain},\text{structural}, \text{none}\}$, representing how notes relate to one another in the score. Structural edges connect notes that are connected with Schenkerian prolongations. Note that since there can only be one edge type between any two nodes, we must choose a precedence order for edge types that might coexist. For instance, if voice edges are overwritten, the music cannot be reconstructed. Most \textit{Surface level} edges (edges that are inherent to the written music such as forward, onset, and sustain) are mutually exclusive, but voice edges are a subset of forward edges, and structural edges often coincide with forward/voice edges. Thus, surface level edges take precedence over any overlapping structural edges and voice and voice edges take precedence over forward edges.

\textbf{\textit{Forward Process:}} A graph $G = (\mathbf{X}, \mathbf{E})$ is comprised of node embedding matrix $\mathbf{X} \in \{0,1\}^{n\times |\mathcal{X}|}$, where each row is a one-hot encoding $\mathbf{x}_i \in \{0,1\}^{|\mathcal{X}|}$ for graph nodes $i = 1,\dots,n$, and edge embedding tensor $\mathbf{E} \in \{0,1\}^{n\times n\times |\mathcal{E}|}$, which describes each edge $\mathbf{e}_{i,j} \in \{0,1\}^{|\mathcal{E}|}$ from node $i$ to node $j$ as a one-hot encoding. Discrete graph diffusion applies noise independently to each node and edge, similar to pixels in image diffusion. At each forward diffusion step $1,\dots,t,\dots,T$, node and edge class transition probability matrices are defined as $\mathbf{Q}^t_{X} \in [0,1]^{|\mathcal{X}|\times|\mathcal{X}|}$ and $\mathbf{Q}^t_{E} \in [0,1]^{|\mathcal{E}|\times|\mathcal{E}|}$ respectively. In both matrices, each row describes the transition probability from category $i$ to all other categories $j$ such that $\sum_j [\mathbf{Q}_{X}^t]_{i,j} = \sum_j[\mathbf{Q}_{E}^t]_{i,j} = 1$ for all $i$. We can then sample each node and edge at time $t$ (forming graph $G^t$) given graph $G^{t-1}$ using the transition probability $q(G^t \mid G^{t-1})$, taken as the product of the node-specific transition probabilities $\mathbf{X}^{t-1} \mathbf{Q}^t_{X}$ and the edge-specific probabilities $\mathbf{E}^{t-1} \mathbf{Q}^t_{E}$. Furthermore, we can determine the distribution at any time directly from the original graph $G^0$ using the well-known Chapman-Kolmogorov equation, notated here as $\prod_{\tau=1}^{t} \mathbf{Q}^\tau_X =: \bar{\mathbf{Q}}^t_X$ and $\prod_{\tau=1}^{t} \mathbf{Q}^\tau_E =: \bar{\mathbf{Q}}^t_E$

\textbf{\textit{Reverse Process:}} The denoising process is estimated using a model $\phi_\theta$ parameterized by $\theta$. This model is trained to directly estimate a graph representing a piece of music $G^0 = (\mathbf{X}^0, \mathbf{E}^0)$ given a noisy graph at any time step $G^t=(\mathbf{X}^t, \mathbf{E}^t)$. We denote the predicted probabilities for each node in the original graph $G^0$ as $\hat{p}_{\mathbf{X}} \in [0,1]^{n\times |\mathcal{X}|}$.

In our implementation, edges are predefined and static based on the rhythmic framework of sampled musical material (Section \ref{subsec:rhythmic_sampling}). This assumption simplifies the diffusion problem considerably, as we are able to set the edge transition matrix to the identity $\mathbf{Q}^t_{E} = \bar{\mathbf{Q}}^t_{E} = \mathbf{I}_{|\mathcal{E}|}$. Following \citep{vignac2023digressdiscretedenoisingdiffusion}, we set $\bar{\mathbf{Q}}^t_\mathbf{X} = \bar{\alpha}^t\mathbf{I}_{|\mathcal{X}|} + (1 - \bar{\alpha}^t)\mathbf{1}[\mathbf{m_X}]'$, where $\mathbf{m_X}$ is the marginal distribution vector for node types, $[\cdot]'$ is the transpose, and $\mathbf{1}$ is a ones vector. Here, $\bar{\alpha}^t = \prod_{\tau=1}^t \alpha^t$ is the noise schedule hyperparameter that goes from 1 to 0 (true data to complete noise) according to the cosine schedule, $[\alpha^t]^2 = f^t/f^0$, where $f^t = \text{cos}(((t/T + s)/(1 + s)) \cdot (\pi/2))^2$, and $s$ is a small number (e.g. 0.008) \citep{nichol2021improveddenoisingdiffusionprobabilistic}. By freezing the edges of the graph, the reverse diffusion objective is simplified considerably. The DiGress loss (\cref{eqn:full_loss} in Appendix A) is reduced to $\mathcal{L}(\hat{p}_{\mathbf{X}}, \mathbf{X}) = \sum_{i=1}^n\text{cross-entropy}(\mathbf{x}_i, [\hat{p}_{\mathbf{X}}]_i)$, only attending to the predictions for nodes. Further, we only require $\hat{p}_{\mathbf{X}}$ to estimate reverse diffusion transitions $p_\theta(G^{t-1}\mid G^t) = \prod_{i=1}^n p_\theta(\mathbf{x}_i^{t-1}\mid \mathbf{x}_i^t)$ (compare with \cref{eqn:graph_transition_prob,eqn:node_transition_prob} in Appendix A).

% The trained denoising model can then be used to sample new graphs, using its predictions $\hat{p}_{\mathbf{X}}$ to estimate reverse diffusion iterations $p_\theta(G^{t-1}\mid G^t) = \prod_{i=1}^n p_\theta(\mathbf{x}_i^{t-1}\mid \mathbf{x}_i^t)$. Each term is computed by marginalizing over network predictions,
% $$p_\theta(\mathbf{x}_i^{t-1}\mid \mathbf{x}_i^t) = \sum_{c=1}^{|\mathcal{X}|}p_\theta\Bigl(\mathbf{x}_i^{t-1}\mid \mathbf{x}_i^0 = \mathbbm{1}_c, \mathbf{x}_i^t\Bigr)\bigl[\hat{p}_\mathbf{X}\bigr]_{i,c}$$
% where $\mathbbm{1}_c$ is the one-hot encoding for class $c$ and we choose
% \begin{gather*}
% p_\theta\Bigl(\mathbf{x}_i^{t-1}\mid \mathbf{x}_i^0 = \mathbbm{1}_c, \mathbf{x}_i^t\Bigr) = \\
% \begin{cases} 
% q(\mathbf{x}_{i}^{t-1} \mid \mathbf{x}^0_i = \mathbbm{1}_c, \mathbf{x}_i^t) & \text{if } q(\mathbf{x}_i^t \mid \mathbf{x}_i^0 = \mathbbm{1}_c) > 0, \\
% 0 & \text{otherwise}.
% \end{cases}
% \end{gather*}

% \begin{algorithm*}
% \caption{Training ProGress}\label{algorithm:training}

% \begin{algorithmic}

% \end{algorithmic}

% \end{algorithm*}

% \begin{algorithm*}
% \caption{Sampling from ProGress}\label{algorithm:sampling}

% \begin{algorithmic}
% \Require Trained SchenkerLink model $\zeta$
% Trained denoising model $\phi_\theta$
% \State Sample rhythmic skeleton $m$ from music dataset
% \State $G_{\text{multi}} \leftarrow \text{graph}(m)$

% \end{algorithmic}

% \end{algorithm*}

The DiGress framework expects nodes with only discrete, one-hot encoded embeddings. However, beyond discrete scale degrees, we include discrete and continuous rhythmic features, bundled in a matrix $\mathbf{R} \in \mathbb{R}^{n\times |\mathcal{R}|}$, where $\mathcal{R}$ represents the set of rhythmic features. Because $\mathbf{R}$ is determined and unchanging from the beginning of the process, it can be incorporated in every denoising iteration to model $\phi_\theta$ (recall Figure \ref{fig:model_overview}). More specifically, the input of the denoising model $\phi_\theta$ should be $G^t = \bigl([\mathbf{X}^t \mid\mid \mathbf{R}], \mathbf{E}\bigr)$, where $[\cdot\mid\mid\cdot]$ denotes column concatenation. When performing the reverse iterations during inference, we implement Stochastic Control Guidance \cite{huang2024symbolicmusicgenerationnondifferentiable} to avoid certain harmonic intervals, undesired contrapuntal motions, and repetitive melodic lines. Depending on the genre, any quantifiable rules may be added to guide the diffusion process.

\subsection{Inference and Phrase Fusion}
\label{subsec:inference}

\textbf{\textit{Music Realization:}} Because we limited the classes of individual nodes to the global scale degrees (e.g. $\hat{1}$, $\hat{4}$, or $\flat\hat{3}$), they cannot be directly interpreted as music. Rather, they must be mapped to specific pitches in specific octaves (e.g. C4, F2, or E$\flat$4). The simplest approach is to follow the path of \textit{smoothest voice leading} for each string of nodes connected by forward edges: i.e., starting in a register common to the dataset, for each scale degree we place it according to the smallest interval between the previous note and itself. However, this approach often leads to melodies that go extremely high or low. 
Instead, we define a central pitch for each voice, which serves as a fall back if a voice gets too far away. If it is possible for consecutive notes to be a step away, they will always follow step-wise motion. If there is a larger interval between consecutive notes, the voice will find the closest note to the central pitch. This approach ensures smooth voice leading is achieved while constraining the voice range.

\begin{wrapfigure}{r}{0.45\textwidth}
\vspace{-0.5cm}
  \begin{center}
    \includegraphics[width=0.43\textwidth]{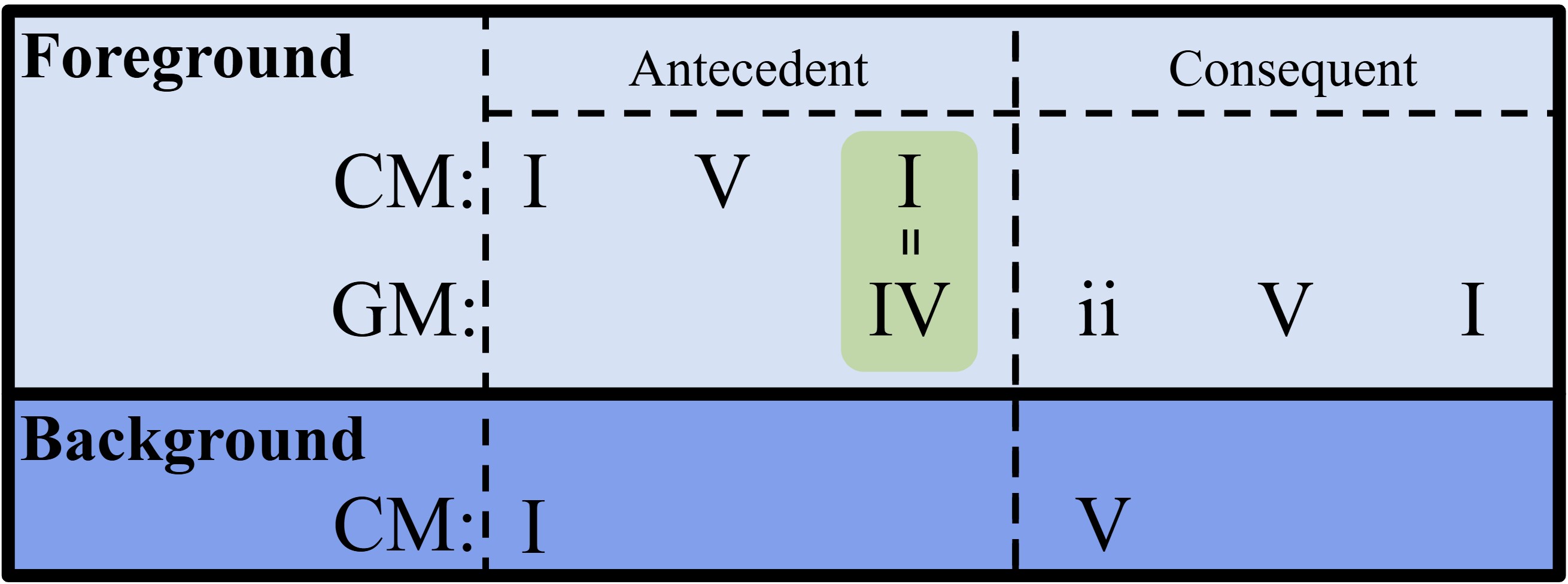}
  \end{center}
  \caption{Example phrase fusion via pivot chord modulation from C Major (CM) to G Major (GM). The light and dark blue represent foreground and background analysis, respectively. The \textit{antecedent} is in CM, leading to GM in the \textit{consequent} by reinterpreting the final antecedent ``I'' as ``IV'' in the new key.}
  \vspace{-0.5cm}
\label{fig:pivot_modulation_example}
\end{wrapfigure}

\textbf{\textit{Rejection Sampling and Musical Analysis:}} Through diffusion, we generate $B$ musical phrases. Once all phrases are generated (which may be done in parallel), we impose a rule-based rejection sampling to discard poor quality musical phrases. Similar to the Stochastic Control Guidance mentioned in Section \ref{subsec:diffusion_modeling}, we reject samples with improper harmonic intervals or contrapuntal motions. Additionally, we analyze the phrase for possible harmonic progressions based on the desired genre. If no harmonic progression can make sense of the phrase, it is discarded. 

During the analysis process, we keep track of possible starting and end harmonies and melodic tones. Because important structural events tend to happen at the beginnings and endings of phrases according to Schenkerian theory \cite{cadwallader1998analysis}, we can fuse phrases together to match a common Schenkerian structure such as the one found in the purple box of Figure \ref{fig:model_overview}. By incorporating such Schenkerian structure, we ensure the generated music has meaningful local and global harmonic variation with direction.

% \begin{figure}
%     \centering
%     \includegraphics[width=0.5\linewidth]{figs/pivot_fusion_example.jpg}
%     \caption{Example phrase fusion via pivot chord modulation from C Major to G Major. The light and dark blue represent foreground and background analysis, respectively. The \textit{antecedent} is primarily in C Major, leading to G Major in the \textit{consequent} by reinterpreting the final ``I'' as a ``IV'' in the new key.}
%     \label{fig:pivot_modulation_example}
% \end{figure}
\begin{figure}[H]
    \centering
    \vspace{-0.2cm}
    \includegraphics[width=0.9\linewidth]{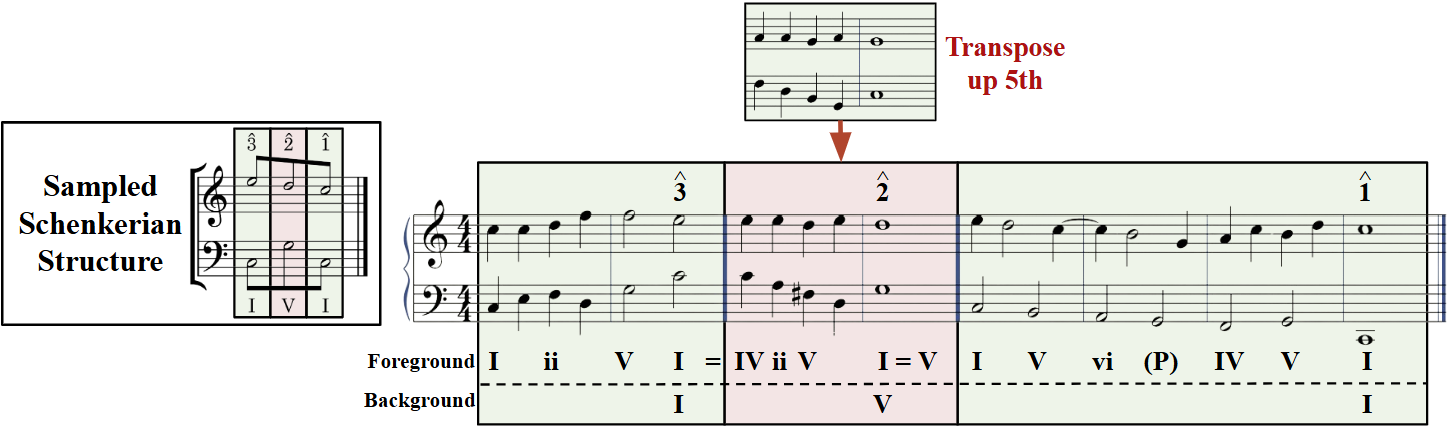}
    \caption{A common Schenkerian structure as three phrases of generated music. Green and red represent music in the home and dominant key, respectively. Here, the 2nd phrase was originally generated in the home key, but is transposed to the dominant via our fusion method in Section \ref{subsec:inference}. }
    \vspace{-0.5cm}
    \label{fig:structure_realization}
\end{figure}

\textbf{\textit{Phrase Fusion:}} To create a smooth transition between phrases, we employ a pivot chord modulation scheme. Say we want to ``modulate'' from the tonic key ``I'' in a \textit{antecedent} phrase to the dominant key ``V'' in a \textit{consequent} phrase (see Figure \ref{fig:pivot_modulation_example} for example). We first assume all phrases are based in a particular key (e.g. C Major/Minor). If the antecedent ends on a tonic ``I'' harmony, the consequent can reinterpret the tonic harmony as a surface level subdominant ``IV'' in the deeper level motion to the dominant ``V.'' Therefore, we search our dictionary of organized phrases for a phrase that begins on a local harmony that typically comes after a ``IV'' harmony. The sampled phrase can then be transposed to the desired key (dominant ``V'' in our example here) and appended to the antecedent as the consequent phrase. Similar transitions can move from one key to any other.

\textbf{\textit{Sampling Schenkerian Structure:}} To determine the overall structure of our generated music, we first gather common Schenkerian structures from the literature. From the set of expert SchAs, we extract the deep middleground structural harmonic progressions and their associated bass and treble notes. 

For instance, the most famous structure in SchA is a 3-line \textit{Ursatz} (depicted in Figure \ref{fig:structure_realization}). The harmonic progression follows a tonic-dominant-tonic ($I-V-I$) structure with root position bass notes and a stepwise descending third in the treble ($\hat{3}-\hat{2}-\hat{1}$). One realization of this structure would involve three phrases. The first phrase would end in the home key with an authentic cadence ($V-I$) and $\hat{3}$ in the treble voice. The second phrase would end with an authentic cadence in the dominant key ($V$) with global $\hat{2}$ (local $\hat{5}$) in the treble voice. Finally, phrase three would end with a perfect authentic cadence in the home key; it would end with $V-I$ in the bass and $\hat{1}$ in the treble.

\section{Experiments}
\label{sec:experiments}

We ran several experiments, including a human survey, ablation studies, and genre flexibility demonstrations. Full survey results, ablation studies, and implementation details can be found in Appendices B--D. Musical samples and genre flexibility demonstrations may be found on our Github page\footnote{https://anonymousforpeerreview.github.io/ProGressDemo/}. Our model was trained on all individual phrases of the Bach chorales that are based in their respective global tonics. We provide the full survey instrument and excerpts in the Supplemental Materials. Reproducible code will be made available pending acceptance.%\footnote{To be added pending acceptance.}. 

% \begin{wrapfigure}{l}{0.42\textwidth}
%   \begin{center}
%     \includegraphics[width=0.40\textwidth]{figs/progress_weird_results_v2.png}
%   \end{center}
%   \caption{``Weird or bad'' survey results.}
%   \label{fig:weird_or_bad}
% \end{wrapfigure}

% \begin{wrapfigure}{r}{0.45\textwidth}
%   \begin{center}
%     \includegraphics[width=0.43\textwidth]{figs/progress_preference_v2.png}
%   \end{center}
%   \caption{Pairwise model preference for all survey participants.}
%   \label{fig:preference}
% \end{wrapfigure}

\begin{figure}[htbp]
\vspace{-0.2cm}
  \centering
  \begin{minipage}{0.48\textwidth}
    \centering
    \includegraphics[width=0.8\textwidth]{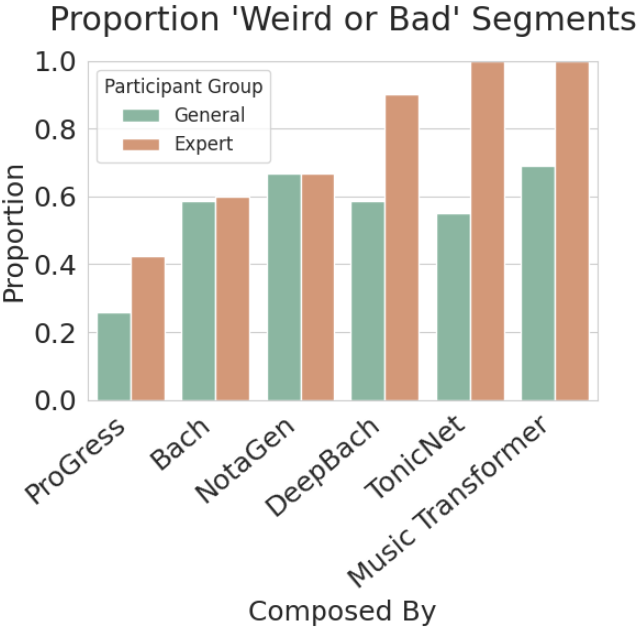}
    \caption{``Weird or bad'' survey results.}
    \label{fig:weird_or_bad}
  \end{minipage}%
  \hfill
  \begin{minipage}{0.48\textwidth}
    \centering
    \includegraphics[width=0.9\textwidth]{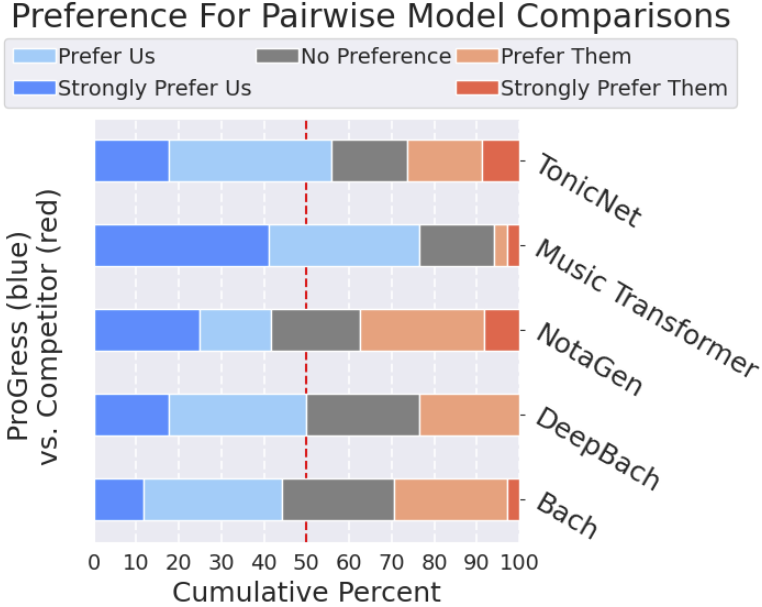}
    \caption{Pairwise model preference for all survey participants.}
    \label{fig:preference}
\end{minipage}
\vspace{-0.2cm}
\end{figure}

\textbf{\textit{Survey Design:}} For our subjective experiments, we use the same survey instrument as \citep{hahn_schenkcomposer} and \citep{hahn2024senthymnent}. We compare against Bach and several models specialized for Bach chorale generation: DeepBach \cite{hadjeres2017deepbach}, NotaGen \cite{wang2025notagenadvancingmusicalitysymbolic}, Music Transformer \cite{huang2018music}, and TonicNet \cite{peracha2020}. For each pair of chorales we asked: \textbf{1)} On a scale of 0 (not enjoyable) to 10 (very enjoyable), how would you rate Chorale $X$? \textbf{2)} On a scale of 0 (certain it's by a computer) to 10 (certain it's by a human), what is your degree of belief that a human composed Chorale $X$? \textbf{3)} Which Chorale do you prefer? (a) strongly prefer 1, (b) prefer 1, (c) no clear preference, (d) prefer 2, (e) strongly prefer 2. \textbf{4)} Were there any parts of Chorale $X$ that stood out as sounding weird or bad to you? (yes=1, no=0).

\textbf{\textit{Results:}} Our final dataset consists of 45 participants. Of those, 13 \textit{expert participants} reported studying music privately for more than 5 years and gave correct answers to skill screening questions. We found that \ourmethod{} outperformed other methods, \textit{and even Bach}, in all qualitative metrics. Observing the ``weird or bad'' question results seen in Figure \ref{fig:weird_or_bad}, we see that \ourmethod{} performs substantially better than other models in both the general and expert participant groups. Bach's score lies comfortably in the middle. We believe this is because \ourmethod{} is more structured than other deep learning models and less harmonically adventurous than Bach.

% \begin{figure}
%     \centering
%     \includegraphics[width=0.47\linewidth]{figs/progress_preference_v2.png}
%     \caption{Pairwise model preference for all survey participants.}
%     \label{fig:preference}
% \end{figure}

In Figure \ref{fig:preference}, we see that participants generally prefer \ourmethod{} over the competitors. NotaGen nearly tied with \ourmethod{}, however our model uses substantially fewer parameters than NotaGen (3 million vs 516 million, respectively). While Bach had around double the proportion of perceived ``weirdness'' in his music, participants did not show a strong preference for our model over Bach. However, we find that \ourmethod{} outperforms Bach in ``enjoyability'' with statistical significance (see Appendix B). 

\section{Conclusion}

We introduce a hybrid approach in which GNNs and music-theoretical structures and principles work together to produce novel, coherent music in various styles. Through our survey experiment, we show that \ourmethod{}'s careful music-hierarchical composition style outperforms the stream-of-consciousness approach of several deep learning models.

\bibliographystyle{plainnat}
\bibliography{bibliography.bib}

\section*{Appendix}

\section*{A. Diffusion Preliminaries and DiGress Details}

\subsection*{Denoising Diffusion Probabilistic Models}

\begin{figure}[H]
    \centering
    \includegraphics[width=0.6\linewidth]{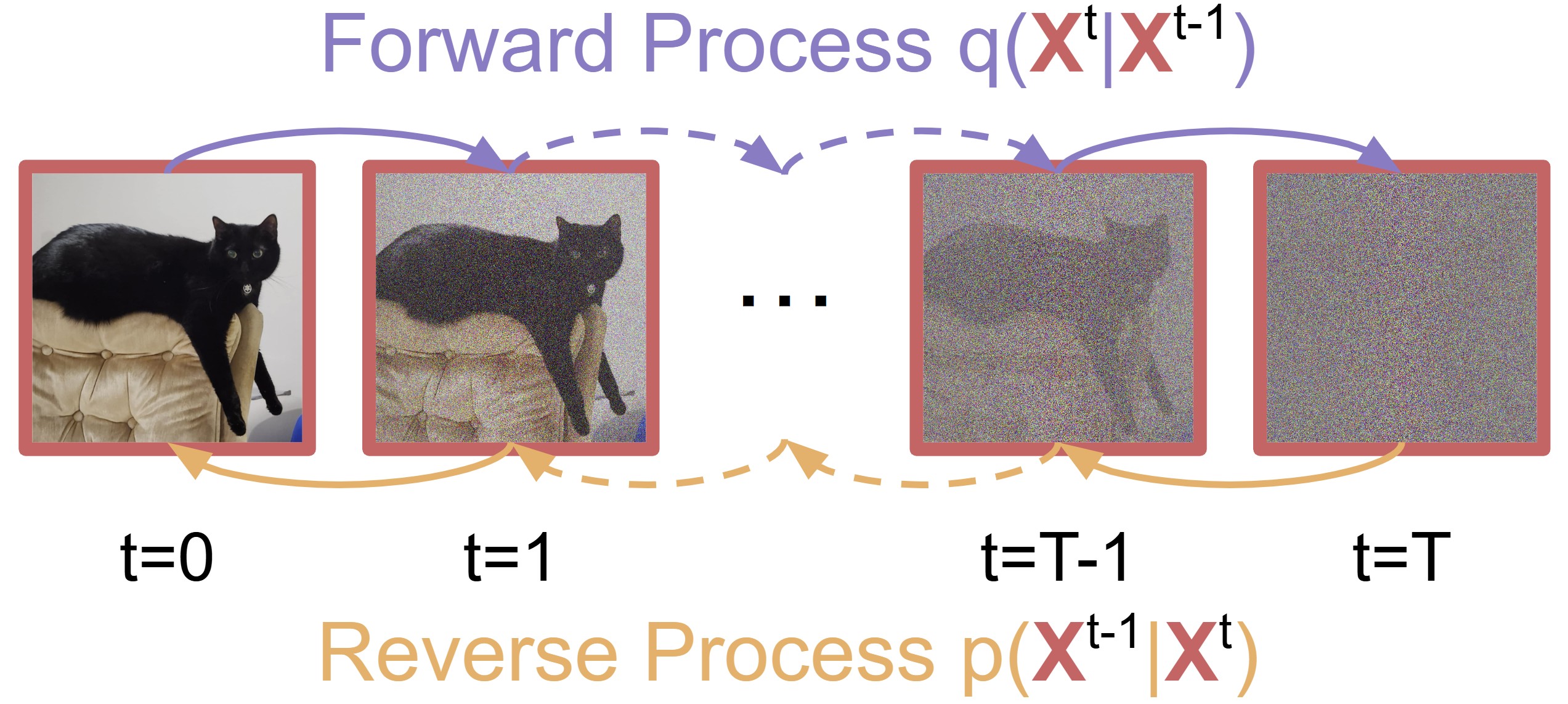}
    \caption{Example of the diffusion process on image data of the first author's cat.}
    \label{fig:image_diffusion}
\end{figure}

Denoising Diffusion Probabilistic Models (DDPMs; introduced by \citep{sohl2015deep}) aim to generate meaningful data (e.g. images or audio) by \textit{denoising} corrupted data. There are two main processes involved in DDPMs that assume a Markov process: the forward (encoding) process $q(\mathbf{X}^{1:T}\mid \mathbf{X}^{0}) = \prod_{t=1}^T q(\mathbf{X}^t\mid\mathbf{X}^{t-1})$, where $\mathbf{X}^t$ is the data after $t=1,\dots,T$ steps of corruption or noise addition, and the reverse (decoding) process $p(\mathbf{X}^{0:T}) = p(\mathbf{X}^T)\prod_{t=1}^T p(\mathbf{X}^{t-1}\mid\mathbf{X}^t)$, which aims to undo the data corruption process or find novel clean data from noise (Figure \ref{fig:image_diffusion}). 
 
Most work in continuous spaces defines the distributions for forward and reverse precesses to be Gaussian \citep{ho2020denoisingdiffusionprobabilisticmodels, song2022denoisingdiffusionimplicitmodels, chen2020wavegradestimatinggradientswaveform, nichol2021improveddenoisingdiffusionprobabilistic}. Even when dealing with categorical data, Gaussian noise is common; categories are treated as one-hot encodings with continuous values \citep{niu2020permutationinvariantgraphgeneration, jo2022scorebasedgenerativemodelinggraphs}. Many works have adapted diffusion for discrete spaces \citep{hoogeboom2021argmaxflowsmultinomialdiffusion,johnson2021inplacecorruptioninsertiondeletion, yang2023diffsounddiscretediffusionmodel,austin2023structureddenoisingdiffusionmodels}. 

%In practice, a denoising model $\phi_\theta$ is trained to handle every iteration of the reverse process by directly estimating the clean data $\mathbf{X}^0$.

Our model follows the setting of \citep{austin2023structureddenoisingdiffusionmodels} and \citep{vignac2023digressdiscretedenoisingdiffusion}, where a data point $\mathbf{x}^0\in \{0,1\}^d$ is a one-hot encoding of $d$ categories and the noise is represented by a series of transition matrices $(\mathbf{Q}^1,\dots,\mathbf{Q}^T)$. These transition matrices are defined such that $[\mathbf{Q}^t]_{i,j}$ represents the probability of moving from state $i$ to state $j$ and $q(\mathbf{x}^t\mid\mathbf{x}^{t-1}) = \mathbf{x}^{t-1}\mathbf{Q}^t \in [0,1]^d$.

\subsection*{Discrete Diffusion with Graph Neural Networks}
\label{subsec:gnn_diffusion_background}

Graphs are a natural medium to represent hierarchy in music \citep{jeong2019graph, ni2024new}. Methods to extract meaningful features from graphs are thus of critical interest, and in the context of deep learning, Graph Neural Networks (GNNs) stand out for their effectiveness. GNNs generalize the discrete convolutions of Convolutional Neural Networks (CNNs) to graphs, where filters perform local neighborhood aggregation over the node space. Following the work of \citep{jeong2019graph, ni2024new}, we consider GNNs that operate over heterogeneous, directed graphs.

\begin{figure}
    \centering
    \includegraphics[width=1.0\linewidth]{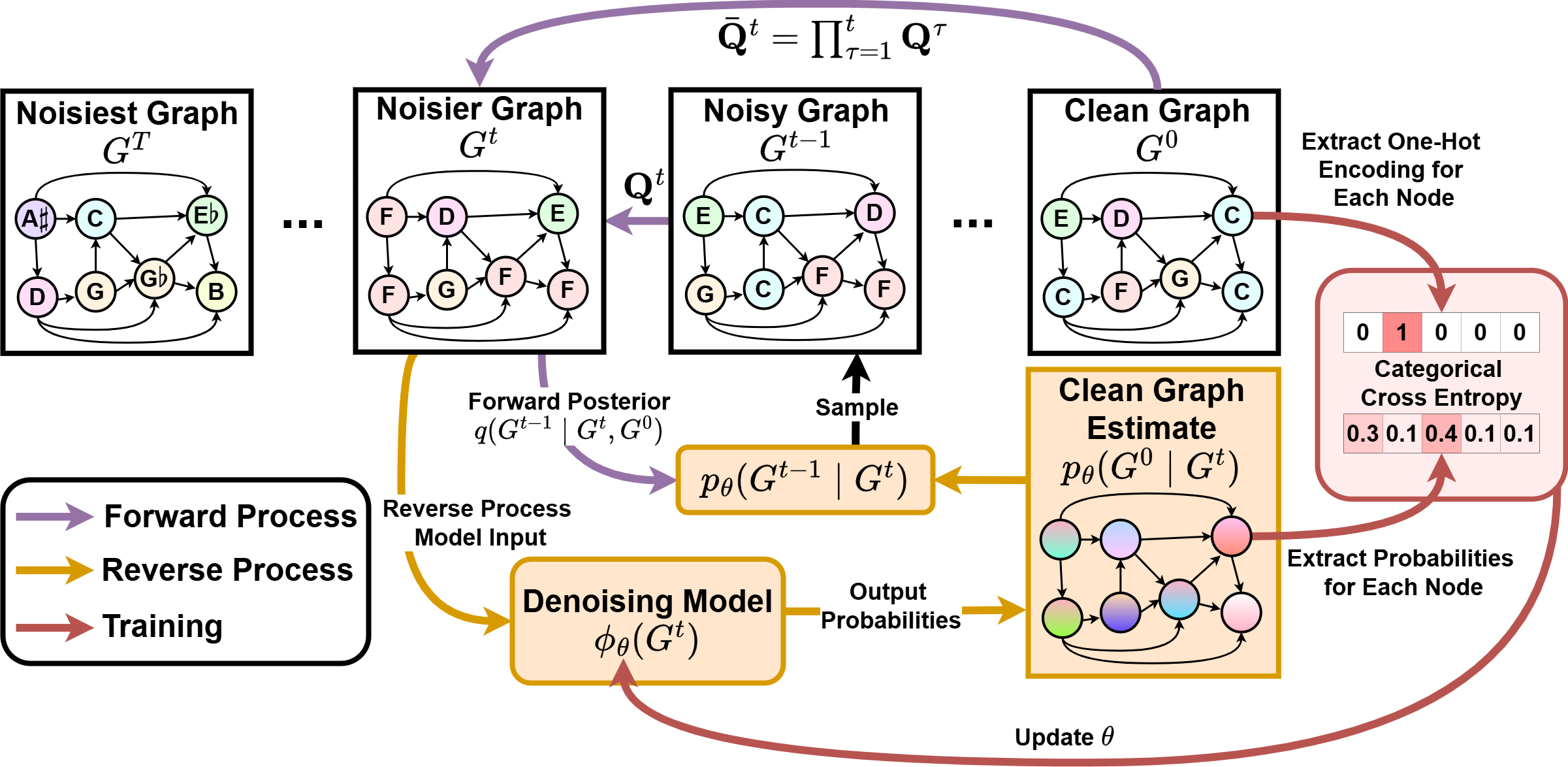}
    \caption{Overview of the DiGress model for our music application adapted from Figure 1 of \cite{vignac2023digressdiscretedenoisingdiffusion}.}
    \label{fig:digress_overview}
\end{figure}

We are particularly interested in the discrete graph diffusion setting introduced by \citep{vignac2023digressdiscretedenoisingdiffusion}, visualized in Figure \ref{fig:digress_overview}. Given a set of node categories $\mathcal{X}$ and edge categories $\mathcal{E}$, a graph $G = (\mathbf{X}, \mathbf{E})$ is comprised of node embedding matrix $\mathbf{X} \in \{0,1\}^{n\times |\mathcal{X}|}$, where each row is a one-hot encoding $\mathbf{x}_i \in \{0,1\}^{|\mathcal{X}|}$ for graph nodes $i = 1,\dots,n$, and edge embedding tensor 
$\mathbf{E} \in \{0,1\}^{n\times n\times |\mathcal{E}|}$, which describes each edge $\mathbf{e}_{i,j} \in \{0,1\}^{|\mathcal{E}|}$ from node $i$ to node $j$ as a one-hot encoding. Note that the absence of an edge or node is represented as a particular class. Thus, all $\mathbf{x}_i$ and $\mathbf{e}_{i,j}$ are non-empty and have one entry indicating its category.

%consists of rows corresponding to a one-hot encoding $\mathbf{x}_i \in \{0,1\}^{|\mathcal{X}|}$ for graph nodes $i = 1,2,\dots,n$. 

\subsubsection*{Forward Process}

Discrete graph diffusion applies noise independently to each node and edge (like pixels in image diffusion). At each forward diffusion step $1,\dots,t,\dots,T$, node and edge class transition probability matrices are defined as $\mathbf{Q}^t_{X} \in [0,1]^{|\mathcal{X}|\times|\mathcal{X}|}$ and $\mathbf{Q}^t_{E} \in [0,1]^{|\mathcal{E}|\times|\mathcal{E}|}$ respectively. In both matrices, each row describes the transition probability from category $i$ to all other categories $j$ such that $\sum_j [\mathbf{Q}_{X}^t]_{i,j} = \sum_j[\mathbf{Q}_{E}^t]_{i,j} = 1$ for all $i$. We can then sample each node and edge at time $t$ (forming graph $G^t$) given graph $G^{t-1}$ using the following categorical distribution:
$$q(G^t \mid G^{t-1}) = \big(\mathbf{X}^{t-1} \mathbf{Q}^t_{X}, \mathbf{E}^{t-1} \mathbf{Q}^t_{E} \big).$$

Furthermore, we can determine the distribution at any time directly from the original graph $G^0$ using the well-known Chapman-Kolmogorov equation:
$$
q(G^t \mid G^0) = \Biggl(\mathbf{X}^0 \prod_{\tau=1}^{t} \mathbf{Q}^\tau_X,\;\; \mathbf{E}^0 \prod_{\tau=1}^{t} \mathbf{Q}^\tau_E \Biggr) =: \Bigl(\mathbf{X}^0\; \bar{\mathbf{Q}}^t_X, \mathbf{E}^0\; \bar{\mathbf{Q}}^t_E \Bigr).
$$

\subsubsection*{Reverse Process}

The denoising process is estimated using a model $\phi_\theta$ parameterized by $\theta$. This model is trained to directly estimate a graph representing a piece of music $G^0$ given a noisy graph at any time step $G^t$. We denote the predicted probabilities for each node in the original graph $G^0$ as $\hat{p}_{\mathbf{X}} \in [0,1]^{n\times |\mathcal{X}|}$. To avoid clutter, the time superscript $0$ (indicating variables without noise) is implicit in our notation for $\mathbf{X}$, $\mathbf{E}$, $\mathbf{x}$, and $\mathbf{e}$ when no superscript is written. The model is optimized using the cross-entropy loss,
\begin{align}
\mathcal{L}(\hat{p}_{G}, G) =\sum_{i=1}^n\text{cross-entropy}\Bigl(\mathbf{x}_i, \bigl[\hat{p}_{\mathbf{X}}\bigr]_i\Bigr) +
\lambda\sum_{i=1}^n\sum_{j=1}^n \text{cross-entropy}\Bigl(\mathbf{e}_{i,j}, [\hat{p}_\mathbf{E}]_{i,j}\Bigr), \label{eqn:full_loss}
\end{align}

\noindent where $\lambda$ controls the attention balance between edge and node predictions.

The trained denoising model can then be used to sample new graphs, using its predictions $\hat{p}_{\mathbf{X}}$ to estimate reverse diffusion iterations. We model the problem as 
\begin{align}
p_\theta(G^{t-1}| G^t) = \prod_{i=1}^n p_\theta(\mathbf{x}_i^{t-1}| \mathbf{x}_i^t)\prod_{i=1}^n\prod_{j=1}^n p_\theta(\mathbf{e}_{i,j}^{t-1}| G^t). \label{eqn:graph_transition_prob}
\end{align} 

Each term is computed by marginalizing over network predictions,
\begin{align}
p_\theta(\mathbf{x}_i^{t-1}| \mathbf{x}_i^t) = \sum_{c=1}^{|\mathcal{X}|}p_\theta\Bigl(\mathbf{x}_i^{t-1}\mid \mathbf{x}_i^0 = \mathbbm{1}_c, \mathbf{x}_i^t\Bigr)\Bigl[\hat{p}_\mathbf{X}\Bigr]_{i,c} \label{eqn:node_transition_prob}
\end{align} 
where $\mathbbm{1}_c$ is the one-hot encoding for class $c$ and we choose
$$
p_\theta\Bigl(\mathbf{x}_i^{t-1}\mid \mathbf{x}_i^0 = \mathbbm{1}_c, \mathbf{x}_i^t\Bigr) =
\begin{cases} 
q(\mathbf{x}_{i}^{t-1} \mid \mathbf{x}^0_i = \mathbbm{1}_c, \mathbf{x}_i^t) & \text{if } q(\mathbf{x}_i^t \mid \mathbf{x}_i^0 = \mathbbm{1}_c) > 0, \\
0 & \text{otherwise}.
\end{cases}
$$

%We additionally incorporate ``helper'' variables

\noindent Edge transitions are computed in a similar fashion. Graphs are then iteratively sampled using these distributions, where the new graph is used as input for the denoising model $\phi_\theta$ at the next time step.

%: \begin{align}
% p_\theta\bigl(\mathbf{e}_{i,j}^{t-1}\mid \mathbf{e}_{i,j}^t\bigr) = \sum_{c=1}^{|\mathcal{E}|}p_\theta\Bigl(\mathbf{e}_{i,j}^{t-1}\mid \mathbf{e}_{i,j}^0 = \mathbbm{1}_c, \mathbf{e}_{i,j}^t\Bigr)\Bigl[\hat{p}_\mathbf{E}\Bigr]_{i,j,c}. \label{eqn:edge_transition_prob}
% \end{align}

\section*{B. Full Qualitative Survey Results}

Our survey\footnote{Full preview of survey instrument here: https://duke.yul1.qualtrics.com/jfe/preview/previewId/baea6f01-5f30-47b4-b056-f4a9abdb30df/SV\_1zVCXYMgF4KDZS6?Q\_CHL=preview\&Q\_SurveyVersionID=current} begins with the following instructions: ``For the following survey, you will be presented with several pairs of Chorales that aim to imitate the style of J.S. Bach. For each pair of Chorales, you will first be asked to listen to them completely, then answer a series of simple questions. There are 4 total comparisons. Thank you for your time!''

Our survey includes a few screening questions: 1) how often do you listen to music, 2) Have you ever studied music with a private teacher? If so, for how long, 3) What meter best fits [an excerpt of \textit{Ah! Vous dirai-je, maman}], and 4) What is the name of the melodic interval of [two melodic notes]? Self-reported results for experience and skill questions may be found in Table \ref{table:screening}. Skill question responses are divided between a ``nonsense,'' ``wrong,'' and ``correct'' answers, where ``nonsense'' answers use terminology that is not used in music theory. For weekly music listening, 1 reported less than an hour, 33 reported between 1 and 15 hours, and 11 reported more than 15 hours.

\begin{table}[H]
\centering
\caption{Screening question results broken down by reported experience.}
\begin{tabular}{l | ccc | ccc}
\toprule
& \multicolumn{3}{c}{Meter} & \multicolumn{3}{c}{Interval} \\
\cmidrule(lr){2-4} \cmidrule(lr){5-7}
Experience & Nonsense & Wrong & Correct & Nonsense & Wrong & Correct \\
\midrule
0 years       & 4 & 2 & 10 & 6 & 2 & 8 \\
$<$ 5 years   & 0 & 1 & 8  & 1 & 1 & 7 \\
$\geq$ 5 years & 2 & 0 & 18 & 3 & 3 & 14 \\
\bottomrule
\end{tabular}
\label{table:screening}
\end{table}
% \begin{table}[H]
% \centering
% \caption{Screening question results.}
% \begin{tabular}{l | c c c | c c c } 
%         & Meter      &       &       & Interval & & \\
% Experience & Nonsense & Wrong & Correct & Nonsense & Wrong & Correct \\
%  \hline\hline
%  0 years & 4 & 2 & 10 & 6 & 2 & 8\\
%  \hline
%  $<$ 5 years & 0 & 1 & 8 & 1 & 1 & 7\\ 
% \hline
% $\geq$ 5 years & 2 & 0 & 18 & 3 & 3 & 14\\
% \end{tabular}
% \label{table:screening}
% \end{table}

For model comparisons, we note that an official implementation of Music Transformer \cite{huang2018music} is not publicly available, so we trained a model based on https://github.com/gwinndr/MusicTransformer-Pytorch, which has been used for experiments by \citep{huang2024symbolicmusicgenerationnondifferentiable}.

For enjoyability, we compare the mean of each competing excerpt vs$.$ \ourmethod{} using a paired t-test (Table \ref{table:enjoyability}). Similarly, we evaluate mean confidence that each excerpt was composed by a human compared to the actual human-composed excerpt using a paired t-test (Table \ref{table:turing}). We determine binomial confidence intervals for the proportion of participants that strictly preferred \ourmethod{} compared to the competitors, excluding ``no preference'' responses from being counted in favor of \ourmethod{} (Table \ref{table:preference}). Finally, we evaluate whether there is evidence for a difference in the proportion of respondents that identified a ``weird or bad'' sounding excerpt for each competing excerpt vs$.$ \ourmethod{} using a chi-square test.

\begin{table}[H]
\centering
\caption{Proportion of respondents strictly preferring \ourmethod{} (higher is better).}
\begin{tabular}{l c c r} 
Method & Proportion & 95\% CI \\
 \hline\hline
 vs. TonicNet & 0.56 & (0.29, 0.61) \\ 
 vs. Music Transformer & 0.76 & (0.60, 0.88) \\
 vs. DeepBach & 0.50 & (0.34, 0.66) \\
 vs. NotaGen & 0.42 & (0.24, 0.61) \\
 vs. Bach & 0.44 & (0.29, 0.61) \\
\end{tabular}
\label{table:preference}
\end{table}

In Table \ref{table:preference} we show the Clopper-Pearson binomial confidence intervals for the proportion of participants that strictly preferred \ourmethod{} over competitors. Note that we exclude ``no preference'' participants being counted in favor of \ourmethod{}, handicapping our score.

\begin{figure}[H]
    \centering
    \includegraphics[width=0.5\linewidth]{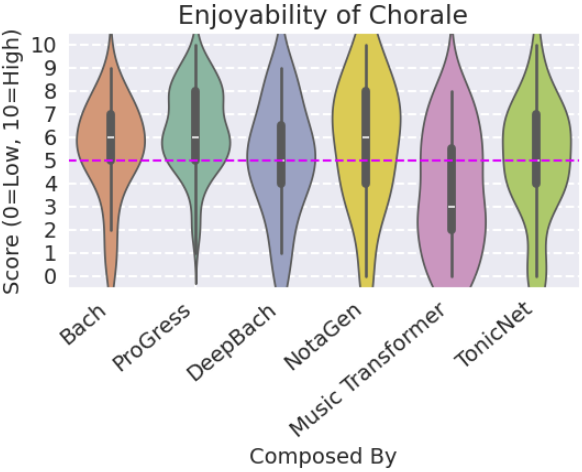}
    \caption{Enjoyability survey results.}
    \label{fig:enjoyability}
\end{figure}

Figure \ref{fig:enjoyability} shows that participants found all models generally enjoyable except Music Transformer. Table \ref{table:enjoyability} shows that within a general population sample, \ourmethod{} is statistically more enjoyable than Bach and all other models except NotaGen.

\begin{table}[H]
\centering
\caption{Enjoyability (higher is better).}
\begin{tabular}{l c c r} 
Method & Mean & 95\% CI & p-value \\ 
 \hline\hline
 \ourmethod{} & 6.37 & (6.08, 6.66) & ref. \\ 
 Bach & 5.47 & (4.80, 6.14) & 0.011 \\
 DeepBach & 5.00 & (4.21, 5.79) & \textless0.001 \\
 NotaGen & 5.75 & (4.59, 6.91) & 0.152 \\
 Music Transformer & 3.68 & (2.81, 4.54) & \textless0.001 \\
 TonicNet & 5.12 & (4.27, 5.96) & 0.001\\
 \end{tabular}
\label{table:enjoyability}
\end{table}

\begin{figure}[H]
    \centering
    \includegraphics[width=0.5\linewidth]{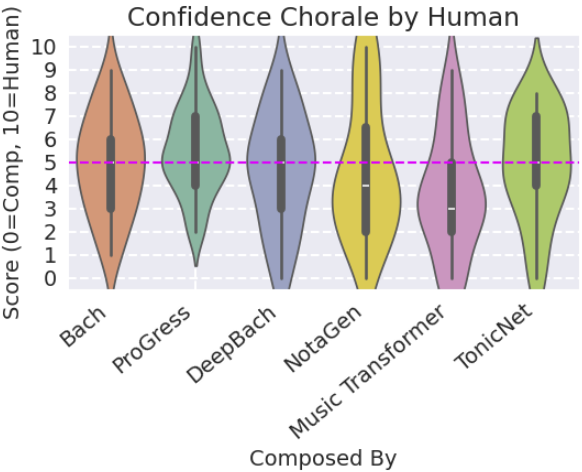}
    \caption{Turing test survey results.}
    \label{fig:turing}
\end{figure}

Figure \ref{fig:turing} and Table \ref{table:turing} show that participants are generally uncertain about whether the excerpts are written by a human or not. Still, \ourmethod{} clearly outperforms other models.

\begin{table}[h!]
\centering
\caption{Confidence of being composed by human (higher is better).}
\begin{tabular}{l c c r} 
Method & Mean & 95\% CI & p-value \\
 \hline\hline
 Bach & 4.76 & (4.06, 5.47) & ref. \\ 
 \ourmethod{} & 5.68 & (5.34, 6.03) & 0.162 \\
 DeepBach & 4.09 & (3.25, 4.92) & 0.212 \\
 NotaGen & 4.63 & (3.38, 5.87) & 0.664\\
 Music Transformer & 2.76 & (1.98, 3.55) & \textless0.001 \\
 TonicNet & 4.06 & (3.21, 4.90) & 0.196\\
\end{tabular}
\label{table:turing}
\end{table}

\section*{C. Ablation Study}

For our ablation studies, we experiment by removing various features within the $\mathbf{R}$ matrix, plus removing the $\mathbf{R}$ matrix altogether. Indeed, we find that the $\mathbf{R}$ matrix is vital to the performance of the network, improving validation loss by approximately 14\%. We report the minimum validation loss for ablated models over 3 runs in Table \ref{table:ablation}.

\begin{table}[h!]
\centering
\caption{Minimum validation loss for ablated models}
\begin{tabular}{l || c | c | c | c | c } 
  & Full & No $\mathbf{R}$ & No metric strength & No duration & No offset\\
 \hline
 Validation Loss & 21.48 & 25.92 & 21.80 & 23.57 & 24.08 \\ 
\end{tabular}
\label{table:ablation}
\end{table}

We also experiment ablating the stochastic control guidance during diffusion inference. Unfortunately, rule guidance did not significantly improve our strict rule-based rejection rate when applied to Bach chorales. The rate when generating 40 samples with and without rule guidance went from 75\% to 77.5\% respectively (lower is better). We hypothesize that larger improvements may be accomplished in other genres with more flexible rules, but leave this to future work.

\section*{D. Implementation Details}

Our model code is in available on Github\footnote{https://github.com/stephenHahn88/ProGress\_Supplement}. In our experiments, our denoising diffusion model consisted of 4 convolutional layers with hidden dimension 256, 8 attention heads, and ran through 100 diffusion steps. It was trained for up to 150 epochs with a batch size 8, using the Adam optimizer. We used a training/validation split of 90/10. These hyperparameters were chosen based on empirical performance on the Bach chorales. We used a single RTX 3060 6gb GPU, which was able to train a full \ourmethod{} model in approximately 47 minutes.

For inference, we generated several hundred phrases and rejected samples that did not follow strict contrapuntal rules based on music theoretical principles of Bach's time. This process took under a minute. These rules included avoiding parallel 5ths and 8ves, avoiding dissonant harmonic intervals (2nds and 4ths) on strong beats, and avoiding improbable harmonic progressions (e.g. V -> IV). These rules may be loosened or adapted for various genres. 

\end{document}